\documentstyle[12pt,psfig,here]{article}

\newcommand{\be}{\begin{equation}}
\newcommand{\ee}{\end{equation}}
\newcommand{\bea}{\begin{eqnarray}}
\newcommand{\eea}{\end{eqnarray}}
\newcommand{\bean}{\begin{eqnarray*}}
\newcommand{\eean}{\end{eqnarray*}}
\newcommand{\ba}{\begin{array}}
\newcommand{\ea}{\end{array}}

\newcommand{\slashl}[1]{\not{\!\!#1}}
\newcommand{\slashs}[1]{\not{\!#1}}
\newcommand{\norsl}{\normalsize\sl}
\newcommand{\norsc}{\normalsize\sc}

\begin{document}
\begin{titlepage}

\title{{ \bf Renormalization of gauge-invariant operators for
	 the structure function \\$g_{2}(x, Q^{2})$}} 

\author{
\norsc  Jiro KODAIRA\thanks{Supported in part by
          the Monbusho Grant-in-Aid for Scientific Research
          No. C-09640364.},
       Takashi NASUNO and  Hiroshi TOCHIMURA\\
\norsl  Dept. of Physics, Hiroshima University\\
\norsl  Higashi-Hiroshima 739-8526, JAPAN\\
\\
\norsc  Kazuhiro TANAKA\thanks{Supported in part by
          the Monbusho Grant-in-Aid for Scientific Research
          No. A-09740215.}\\
\norsl  Dept. of Physics, Juntendo University\\
\norsl  Inba-gun, Chiba 270-16, JAPAN\\
\\
\norsc  Yoshiaki YASUI\\
\norsl  RIKEN BNL Research Center\\
\norsl  Brookhaven National Laboratory\\
\norsl  Upton, NY, 11973, USA\\}

\date{}
\maketitle

\begin{abstract}
{\normalsize
\noindent
We investigate the nucleon's transverse spin-dependent structure
function $g_{2}(x, Q^{2})$ in the framework of the operator product
expansion and the renormalization group.
We construct the complete set of the twist-3 operators
for the flavor singlet channel, and give the relations among them.
We develop an efficient, covariant approach to derive the anomalous
dimension matrix of the twist-3 singlet operators
by computing the off-shell Green's functions.
As an application, we investigate the renormalization mixing 
for the lowest moment case, including the operators proportional
to the equations of motion as well as the ``alien'' operators which are not
gauge-invariant.
}
\end{abstract}

\begin{picture}(5,2)(-340,-620)
\put(2.3,-65){HUPD-9724}
\put(2.3,-80){JUPD-9732}
\end{picture}

\vspace{0.5cm}
\leftline{\hspace{0.5cm}hep-ph/9712395}

\leftline{\hspace{0.5cm}December, 1997}

\thispagestyle{empty}
\end{titlepage}
\setcounter{page}{1}
\baselineskip 24pt

Recently, the nucleon's transverse spin-dependent structure function
$g_{2}(x, Q^{2})$ has been observed\cite{S143} by measuring 
asymmetry in the deep inelastic scattering 
using the transversely polarized target.
The structure function $g_{2}(x, Q^{2})$ plays a
distinguished role in spin physics because
it contains information inaccessible by the more familiar
spin structure function $g_{1}(x, Q^{2})$\cite{SV,JA}: 
$g_{2}(x, Q^{2})$ is related to the nucleon's transverse polarization 
and to the twist-3 operators describing the quark-gluon and
three gluon correlations in the nucleon.

In the framework of the operator product expansion, not only the
twist-2 operators but also the twist-3 operators contribute to
$g_{2}(x, Q^{2})$ in the leading order of $1/Q^{2}$\cite{SV,JA}. 
The $Q^{2}$-evolution of $g_{2}(x, Q^{2})$ is governed by the
anomalous dimensions, which are determined by the renormalization of
these operators. A characteristic feature of the higher twist operators is the
occurrence of the complicated operator mixing under the renormalization:
Many gauge-invariant operators, the number of which increases with spin
(moment of the structure function), 
mix with each other\footnote{Recently, it has been proved
that the twist-3 non-singlet structure functions obey simple DGLAP
evolution equation for $N_{c} \rightarrow \infty$\cite{ABH}.}.
Furthermore, the operators which are proportional to the equations of
motion (``EOM operators''), as well as the ones which are  
gauge-variant (``alien operators''), also participate in the mixing\cite{JCOL1}.

There have been a lot of works on the $Q^{2}$-evolution of
$g_{2}(x,Q^{2})$. Most of them discussed the flavor non-singlet 
case\cite{KOD1,KOD2}. Only a few works treated the singlet case\cite{BKL}:
Bukhvostov, Kuraev and Lipatov derived evolution equations
for the twist-3 quasi-partonic operators.
Recently, M\"uller computed evolution kernel based on the
nonlocal light-ray operator technique, and obtained the 
identical results. However, both of these two works employ a similar
framework based on the renormalization of the nonlocal operators
in the (light-like) axial gauge. Balitsky and Braun
also treated the nonlocal operators although they employed the
background field method. On the other hand, a covariant
approach based on the local composite operators
is missing. Furthermore, some subtle infrared problem occurring in the
renormalization of the generic flavor singlet operators has been emphasized
in Ref.\cite{JCOL2}. Therefore, the computation of the anomalous
dimensions in a covariant gauge in a fully consistent scheme is
desirable and should provide a useful framework.

We develop a covariant framework to
investigate the flavor singlet part of $g_{2}(x, Q^{2})$
based on the operator product expansion and the renormalization group, 
by extending our recent work on the flavor non-singlet part\cite{KOD1,KOD2}.
The purpose of this letter is twofold. Firstly, we provide all
necessary stuffs and convenient techniques:
We list all relevant twist-3 flavor singlet operators
appearing in QCD.
We discuss the relations satisfied by these operators\cite{BKL}.
In particular, we obtain a new operator identity,
which relates the gluon bilinear operator with the trilinear ones.
Based on these developments, we give a basis of independent
operators for the renormalization. We describe a general framework to perform
the renormalization of the twist-3 flavor singlet operators
in a covariant gauge by computing the off-shell Green's functions.
An advantage of our approach is that an infrared cut-off is provided by the
off-shellness of the external momenta.
This allows us to assess unambiguously the renormalization
constant ; otherwise the calculation might
be subtle and potentially dangerous\cite{JCOL2}.
A convenient projection technique\cite{KOD2,TK}
is also introduced to simplify the calculation.
Secondly, we apply our framework to the lowest ($n=3$) moment case,
and demonstrate the consistency and the efficiency of our method:
We investigate in detail the mixing structure of the higher
twist singlet operators under the renormalization.
The EOM operators as well as the alien operators
should be included as independent operators.
We also clarify the connection of our approach with other
works\cite{BKL} based on the on-shell calculation without the EOM operators
nor the alien operators. As a byproduct,
we confirm the prediction of the $Q^{2}$-evolution
obtained in previous works\cite{SV,BKL},
which will be subject to future experimental studies. 

We follow the convention of Refs.\cite{KOD1,KOD2}.
We list the twist-3 flavor singlet operators which contribute
to the moment $\int_{0}^{1}dx x^{n-1}g_{2}(x, Q^{2})\;\;$
$(n = 3, 5, 7, \cdots)$. Firstly, there is a set of 
operators which are bilinear in the
quark fields. These operators can be obtained from the non-singlet
operators, eqs.$(1)$-$(4)$ of Ref.\cite{KOD2} by removing the 
flavor matrices $\lambda_{i}$.
An independent basis can be chosen as (see Refs.\cite{KOD2,KNTTY}):
\bea
  R_{n,l}^{\sigma \mu_{1} \cdots \mu_{n-1}} &=& 
     \frac{1}{2n}\left(V_{l}-V_{n-1-l} + U_{l} + U_{n-1-l} \right)
     \;\;\;\;\;\;\;(l = 1, \cdots, n-2) \ , \label{eq:rl}\\
  R_{n,m}^{\sigma \mu_{1} \cdots \mu_{n-1}}&=&i^{n-2}{\cal S} 
      m \overline{\psi}\gamma_{5}\gamma^{\sigma}
           D^{\mu_{1}}\cdots D^{\mu_{n-2}}\gamma^{\mu_{n-1}}\psi
             - {\rm traces} \ , \label{eq:rm}\\
  R_{n,E}^{\sigma \mu_{1} \cdots \mu_{n-1}} &=& i^{n-2}\frac{n-1}{2n}
   {\cal S} \left[ \bar{\psi}(i \slashl{D} - m ) 
  \gamma_5 \gamma^{\sigma} D^{\mu_{1}} \cdots D^{\mu_{n-2}}
        \gamma^{\mu_{n-1}} \psi \right.\nonumber \\
     & & \qquad + \,\left.\bar{\psi} \gamma_5 \gamma^{\sigma}
    D^{\mu_{1}} \cdots D^{\mu_{n-2}}\gamma^{\mu_{n-1}}(i\slashl{D}
      - m ) \psi \right] - {\rm traces} \ , \label{eq211}
\eea
where
\bean
   V_{l}&=&i^{n} g {\cal S} \bar{\psi}\gamma_{5} 
       D^{\mu_1} \cdots G^{\sigma \mu_l} \cdots
       D^{\mu_{n-2}}\gamma^{\mu_{n-1}} \psi - {\rm traces},\\
   U_{l}&=&i^{n-3} g {\cal S}\bar{\psi}
      D^{\mu_1} \cdots \tilde{G}^{\sigma \mu_l} \cdots
      D^{\mu_{n-2}}\gamma^{\mu_{n-1}} \psi - {\rm traces}.
\eean
In the above equations, (- traces) stands for the subtraction of
the trace terms to make the operators traceless and $D^{\mu}$ is
the covariant derivative.
${\cal S}$ means symmetrization over $\mu_i$ and $g$ the QCD
coupling constant. $m$ represents the singlet component of the quark
mass matrix. 
The operators in eq.(\ref{eq:rl}) contain
the gluon field strength $G_{\mu\nu}$ and the dual tensor
$\widetilde{G}_{\mu \nu}={1\over 2}\varepsilon_{\mu\nu\alpha\beta}
G^{\alpha\beta}$ explicitly.

Secondly, we have the following ``new'' operators (see e.g. Ref.\cite{KNTTY}):
\bea
    T_{n,G}^{ \sigma \mu_{1} \cdots \mu_{n-1} }
       & = &
      i^{ n-1 } \mbox{$ \cal{S} \mit $} \mbox{ $ \cal{A} \mit $ }
      \mbox{$ \cal{S} \mit $} \left[ \widetilde{G}^{ \nu \mu_{1} }
     D^{ \mu _{2} } \cdots D^{ \mu _{n-1} } G_{ \nu }^{ ~ \sigma }
       \right] - \mbox{traces}, \label{eqn:1} \\
    T_{n,l}^{ \sigma \mu_{1} \cdots \mu_{n-1} }
       & = &
      i^{n-2} g \mbox{ $ \cal{ S } \mit $ } \left[
     {G}^{\nu \mu _{1}} D^{ \mu _{2} } \cdots \widetilde{G}^{ \sigma \mu _{l}}
      \cdots D^{ \mu _{n-2} } G_{ \nu }^{ ~ \mu _{n-1} } \right]
          - \mbox{traces} \label{eqn:2}\\
       & & \qquad\qquad\qquad\qquad (l=2,\ldots,\frac{n-1}{2}), \nonumber\\
   T_{n,B}^{ \sigma \mu_{1} \cdots \mu_{n-1} }
      & = &
     i^{n-1} {\cal S}\{ \widetilde{G}^{ \sigma \mu_{1} } D^{ \mu _{2} }
     \cdots D^{ \mu _{n-2} } \}^{a} \{
      - \frac{1}{ \alpha } \partial ^{ \mu_{n-1} }(  \partial ^{ \nu }
      A^{a}_{ \nu } ) +
      g f^{abc}(  \partial ^{ \mu_{n-1}} \overline{\chi}^{ b } ) \chi ^{c}
     \}  \label{eqn:3}\\
     & & \qquad\qquad\qquad\qquad  - \mbox{traces},\nonumber\\
   T_{n,E}^{ \sigma \mu_{1} \cdots \mu_{n-1} }
      & = &
       i^{n-1} {\cal S}\{ \widetilde{G}^{ \sigma \mu_{1} } D^{ \mu _{2} }
        \cdots D^{ \mu _{n-2} } \}^{a} \{
       \left ( D^{ \nu }G_{ \nu }^{~ \mu_{n-1} } \right )^{a} 
                                   \nonumber \\
     & & + \ g \overline{ \psi } t^{a} \gamma^{ \mu_{n-1} } \psi
         + \frac{1}{ \alpha } \partial ^{ \mu_{n-1} }(  \partial ^{ \nu }
            A^{a}_{ \nu }) -
        g f^{abc}(  \partial ^{ \mu_{n-1}} \overline{ \chi } ^{ b } ) \chi ^{c}
              \}  -\mbox{traces}. \label{eqn:4}
\eea
Here $\cal{A}\mit$ antisymmetrizes $\sigma$ and $\mu_{1}$.
The gluon field $A_{ \mu }$ and the covariant derivative 
$ D^{ \mu } $ are in the adjoint representation.
$\chi$ is the ghost field.
$t^{a}$ is the color matrix as $[t^{a}, t^{b}]=i f^{abc} t^{c}$,
${\rm Tr} \left( t^{a} t^{b} \right) = \frac{1}{2} \delta^{ab}$, and 
$\alpha$ is the gauge parameter.
$T_{n,l}$ is trilinear in the gluon field strength $G_{ \mu \nu }$ and 
the dual tensor, and thus represents the effect of the three
gluon correlations. It satisfies the symmetry relation
$T_{n,l}^{ \sigma \mu_{1} \cdots \mu_{n-1} }
= T_{n,n-l}^{ \sigma \mu_{1} \cdots \mu_{n-1} }$.
$T_{n,E}$ is the gluon EOM operator;  it vanishes by the use of
the equations of motion for the gluon,
although it is in general a nonzero operator
due to quantum effects. $T_{n,B}$ is the BRST-exact alien
operator\cite{JCOL1} which is the BRST variation of the operator:
$i^{n-1} {\cal S}\{ \widetilde{G}^{ \sigma \mu_{1} } D^{ \mu _{2} }
\cdots D^{ \mu _{n-2} } \}^{a} \partial^{\mu_{n-1}}\overline{\chi}^{a}
- {\rm traces}$.

We note that the gluon bilinear operator eq.(\ref{eqn:1})
is related to the trilinear ones eq.(\ref{eqn:2}) by
\bea
  T_{n,G}^{ \sigma \mu_{1} \cdots \mu_{n-1}}  &=&
    \sum^{ \frac{n-1}{2} }_{l = 2 } \Biggl[ 
     \frac{( l - 2 )C^{n -2}_{l}}{2(n-1)} -
           \frac{( n - l - 2 )C^{n-2}_{n-l}}{2(n-1)} \nonumber \\
      & & \qquad\qquad  -\ \frac{n(n-2l)C^{n-2}_{l-1}}{2(n-2)(n-1)}
              + (-1)^{l+1} \Biggr]  
         (-1)^{l}T_{n,l}^{ \sigma \mu_{1} \cdots \mu_{n-1}} \nonumber \\
      &+&  \frac{1}{n-1} \left\{
         T_{n,E}^{ \sigma \mu_{1} \cdots \mu_{n-1}}
          + T_{n,B}^{ \sigma \mu_{1} \cdots \mu_{n-1}}
          + \sum^{n-2}_{l = 1} n C^{n - 3}_{l-1} (-1)^{l+1} 
           R^{ \sigma \mu_{1} \cdots \mu_{n-1}}_{n,l}\right\}, \label{eq:oi}
\eea
where $C_{r}^{n}=n!/[ r!(n - r)! ]$. To derive this relation,
we have used $[D_{\mu}, D_{\nu}] = -ig G_{\mu \nu}$ and 
\[ D_{\sigma}G_{\nu \alpha} + D_{\nu} G_{\alpha \sigma}
+ D_{\alpha} G_{\sigma \nu} = 0; \;\;
D_{\sigma}\widetilde{G}_{\nu \alpha} + D_{\nu}
\widetilde{G}_{\alpha \sigma}
+ D_{\alpha} \widetilde{G}_{\sigma \nu} =
\varepsilon_{\nu \alpha \sigma \rho} D_{\lambda} G^{\lambda \rho}, \]
where the first identity is the usual Bianchi identity
while the second one is a consequence of the relation
$g_{\mu \nu} \varepsilon_{\alpha \beta \gamma \delta}
= g_{\mu \alpha} \varepsilon_{\nu \beta \gamma \delta}
+ g_{\mu \beta} \varepsilon_{\alpha \nu \gamma \delta}
+ g_{\mu \gamma} \varepsilon_{\alpha \beta \nu \delta}
+ g_{\mu \delta} \varepsilon_{\alpha \beta \gamma \nu}$.
The operator identity eq.(\ref{eq:oi}) is new,
and is one of the main results of this work.
As a result of eq.(\ref{eq:oi}), 
we can conveniently choose a set of independent operators as
eqs.(\ref{eqn:2})-(\ref{eqn:4}). For the $n$-th moment,
these $\frac{n + 1}{2}$ independent operators
will mix with each other, and with the $n$ gauge-invariant operators
bilinear in the quark fields discussed above,
under the renormalization. 

After the determination of an independent operator's basis,
the renormalization is in principle straightforward:
We follow the standard method to renormalize
the local composite operators\cite{JCOL1}.
We multiply the operators discussed above by a light-like vector
$\Delta_{\mu_{i}}$ to symmetrize the Lorentz indices and to
eliminate the trace terms:
\[ \Delta_{\mu_{1}}\cdots
    \Delta_{\mu_{n-1}} R_{n,l}^{\sigma\mu_{1}\cdots \mu_{n-1}}
    \equiv \Delta \cdot R^{\sigma}_{n,l},\quad
   \Delta_{\mu_{1}}\cdots
    \Delta_{\mu_{n-1}} T_{n,l}^{\sigma\mu_{1}\cdots \mu_{n-1}}
   \equiv
     \Delta \cdot T^{\sigma}_{n,l}. \]
We then embed the operators
${\cal O}_{j} = \Delta \cdot R^{\sigma}_{n,l},\quad
\Delta \cdot T^{\sigma}_{n,l}$, 
into the three-point function as,
\[ \langle 0|{\rm T} {\cal O}_{j}(0) A_{\mu}(x)\psi(y)
    \overline{\psi}(z)|0 \rangle, \quad
   \langle 0|{\rm T} {\cal O}_{j}(0) A_{\mu}(x) A_{\nu}(y)
    A_{\rho}(z)|0 \rangle, \]
and compute the one-loop corrections. We employ the Feynman gauge
($\alpha = 1$)  and use the dimensional regularization.
To perform the renormalization in a consistent manner without subtle
infrared singularities\cite{JCOL2}, we keep the quark and gluon external
lines off-shell; in this case the EOM operators as well as the BRST-exact
operators mix through renormalization as nonzero operators.

One serious problem in the calculation is the mixing
of many gauge-variant as well as BRST-variant EOM operators. As explained
in Refs.\cite{KOD1,KA}, the gauge-variant quark EOM
operators are given by replacing some of the covariant derivatives $D^{\mu_i}$
by the ordinary derivatives $\partial^{\mu_i}$ in 
the corresponding gauge-invariant operator
$R_{n,E}^{\sigma\mu_{1}\cdots \mu_{n-1}}$.
In the present case, the BRST-variant EOM operators obtained similarly
from the gluon EOM operator eq.(\ref{eqn:4}) also participate in the mixing.
However, the problem can be overcome by directly generalizing of the
method employed in Refs.\cite{KOD2,TK}.
We introduce the vector $\Omega_{r}^{ \mu }$
($r= 1, 2, 3$) satisfying  $\Delta_{ \mu }
\Omega_{r}^{ \mu }=0$ for each external gluon line, and contract the
Green's functions as $\Omega_{1}^{\mu}
\langle 0|{\rm T} {\cal O}_{j} A_{\mu} \psi \overline{\psi} |0 \rangle$,
$\Omega_{1}^{ \mu }\Omega_{2}^{ \nu }\Omega_{3}^{ \rho }
\langle 0 | {\rm T} {\cal O}_{j}
        A_{ \mu } A_{ \nu } A_{ \rho } | 0 \rangle$ .
This brings us three merits: Firstly, the tree vertices of the gauge
(BRST) invariant and the gauge (BRST) variant
EOM operators coincide. Thus, essentially only one quark (gluon) EOM
operator is now involved in the operator mixing.
Secondly, the structure of the vertices for the twist-3 operators
is simplified extremely and the computation becomes more tractable
(see eqs.(\ref{eq:3pv1})-(\ref{eq:3pv4}) below).
Thirdly, the three-gluon vertex of the BRST-exact operator eq.(\ref{eqn:3})
vanishes for $\Omega^{\mu}_{2} = \Omega^{\mu}_{3}$
and $\Omega_{1} \cdot \Omega_{2}=0$. Thus, we can exclude the BRST-exact operator
from the gluon three-point functions and compute its mixing separately. 

We now apply our framework to the lowest $(n=3)$ moment.
In this case, there exists no three-gluon operator $\Delta \cdot
T_{3,l}^{\sigma}$. From the one-loop calculation of the
one-particle-irreducible (1PI) three-point function using the
projection by $\Omega_{\mu}^{r}$ (see Fig.1),
it turns out that the renormalization takes the following form:

\bea
\left(
\begin{array}{l}
\Delta\cdot R_{3,1}^{\sigma}\\
\Delta\cdot R_{3,m}^{\sigma} \\
\Delta\cdot T_{3,B}^{\sigma} \\
\Delta\cdot R_{3,E}^{\sigma} \\
\Delta\cdot T_{3,E}^{\sigma}
\end{array}
\right)_{bare}
~=~
\left(
\begin{array}{cccccc}
Z_{11} & Z_{1m} & Z_{1B} & Z_{1E_{F}}&
Z_{1E_{G}} \\
0 & Z_{mm} & 0 & 0 & 0 \\
0 & 0 & Z_{BB} & Z_{BE_{F}} & Z_{BE_{G}} \\
0 & 0 & 0 & Z_{E_{F}E_{F}} & Z_{E_{F}E_{G}} \\
0 & 0 & 0 & Z_{E_{G}E_{F}} & Z_{E_{G}E_{G}}
\end{array}
\right)
\left(
\begin{array}{l}
\Delta\cdot R_{3,1}^{\sigma} \\
\Delta\cdot R_{3,m}^{\sigma} \\
\Delta\cdot T_{3,B}^{\sigma} \\
\Delta\cdot R_{3,E}^{\sigma} \\
\Delta\cdot T_{3,E}^{\sigma}
\end{array}
\right),
\label{eqn:6}
\eea

\vspace{0.3cm}
\noindent
where the operators with (without) the suffix ``$bare$'' are the bare
(renormalized) quantities. The renormalization constant matrix is
triangular; this is consistent with the vanishing physical 
matrix elements of the EOM operators and of the BRST-exact operators
(see below).
\begin{figure}[H]
\begin{center}
\begin{tabular}{ccccccc}
\leavevmode\psfig{file=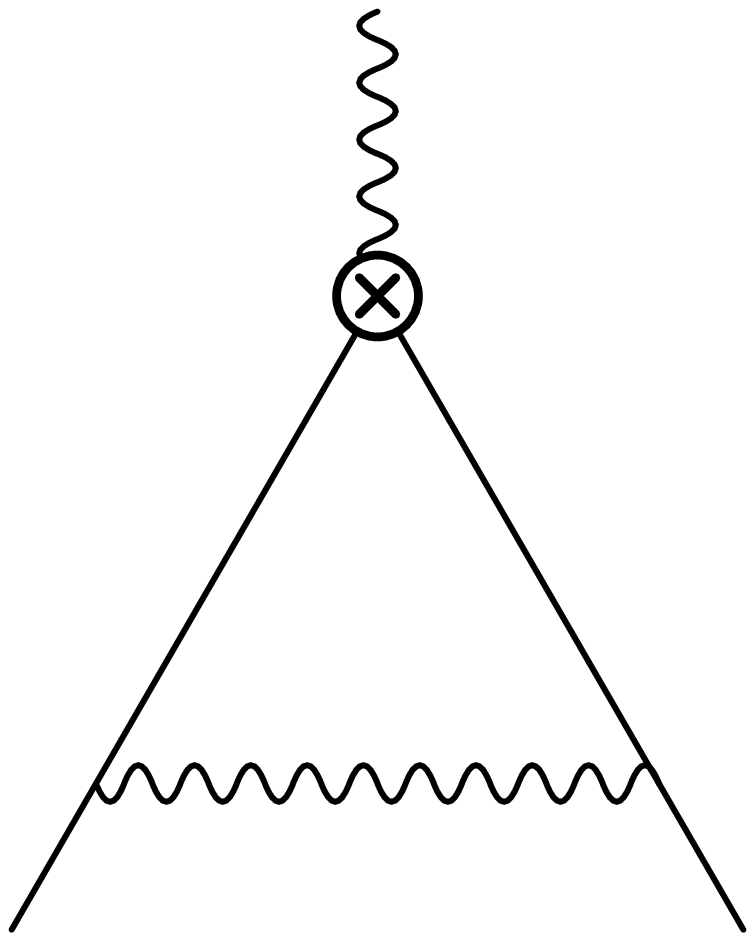,width=1.6cm} &
\leavevmode\psfig{file=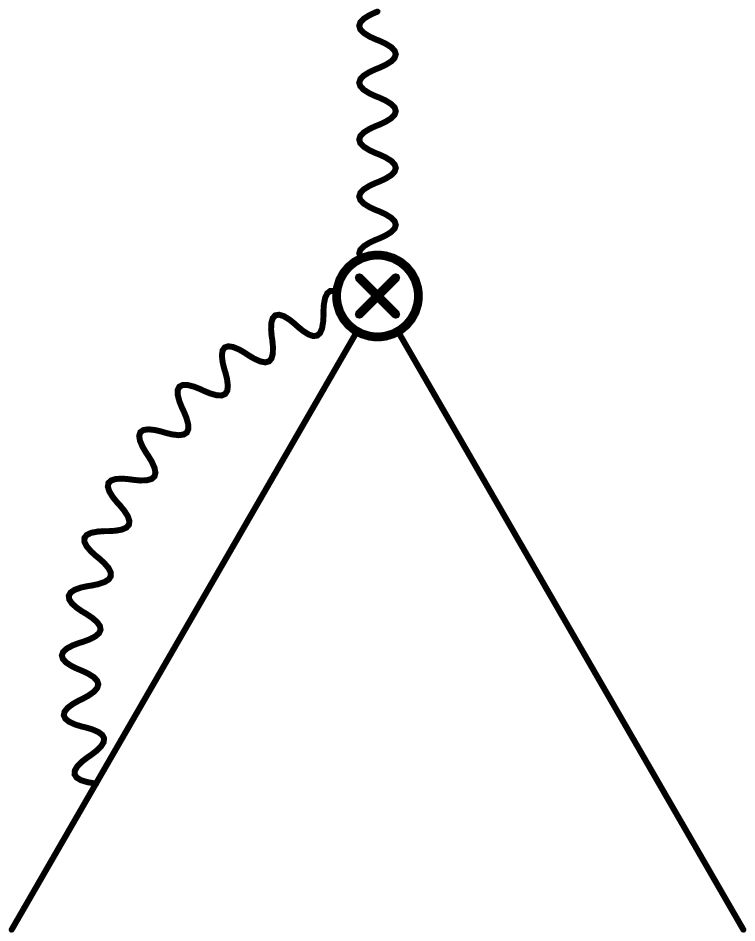,width=1.6cm} &
\leavevmode\psfig{file=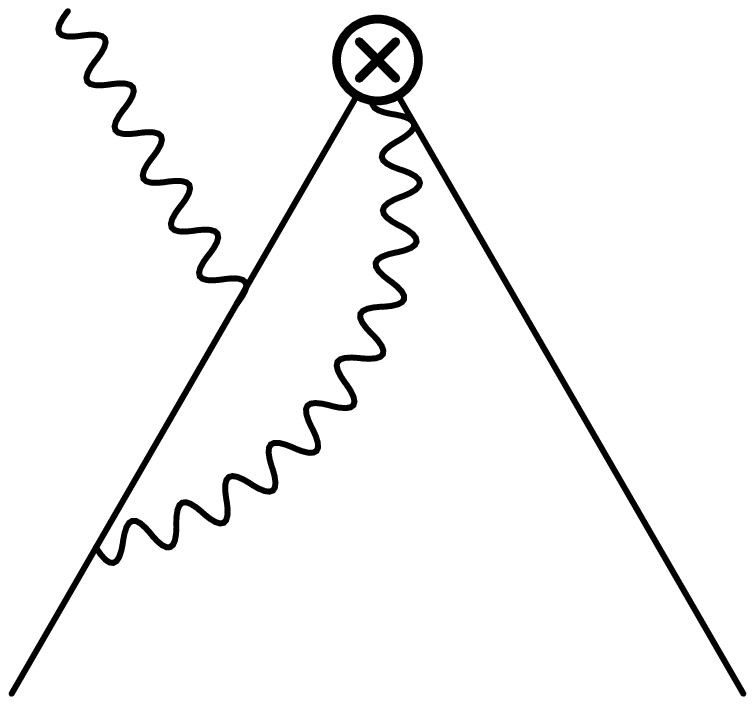,width=1.6cm} &
\leavevmode\psfig{file=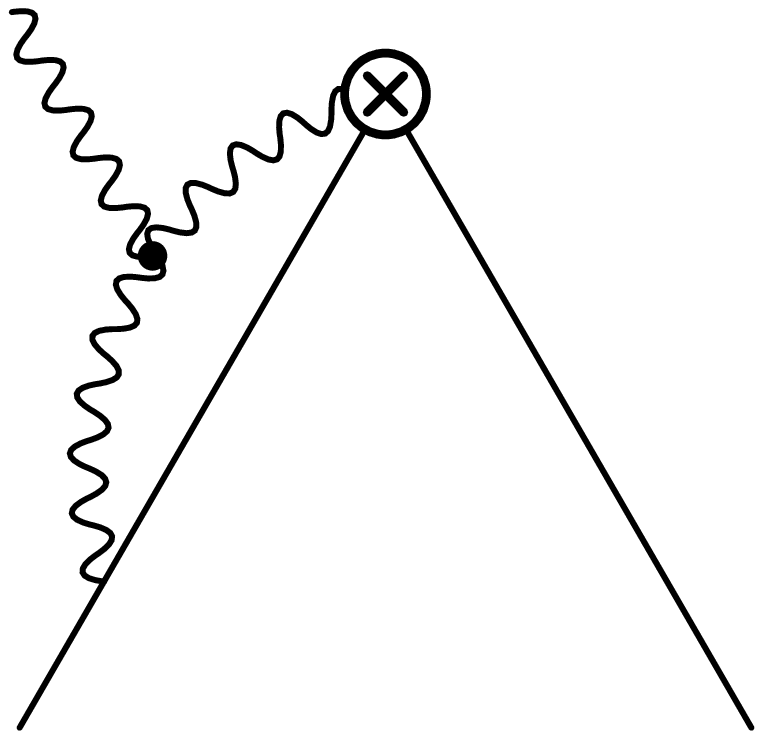,width=1.6cm} &
\leavevmode\psfig{file=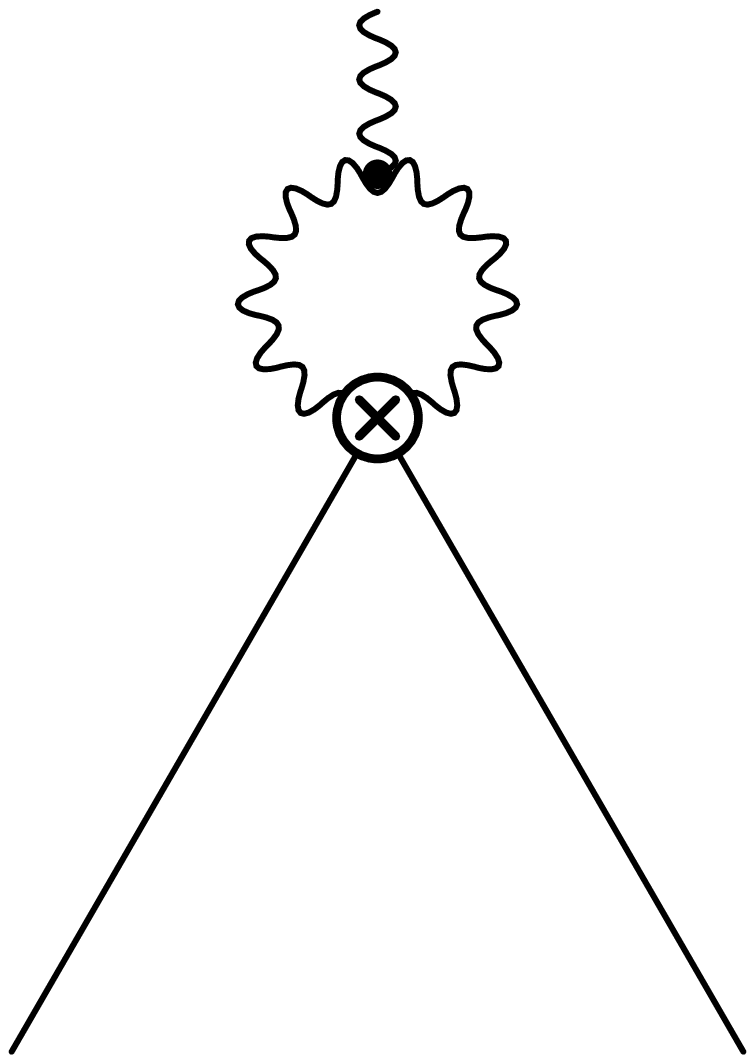,width=1.6cm} &
\leavevmode\psfig{file=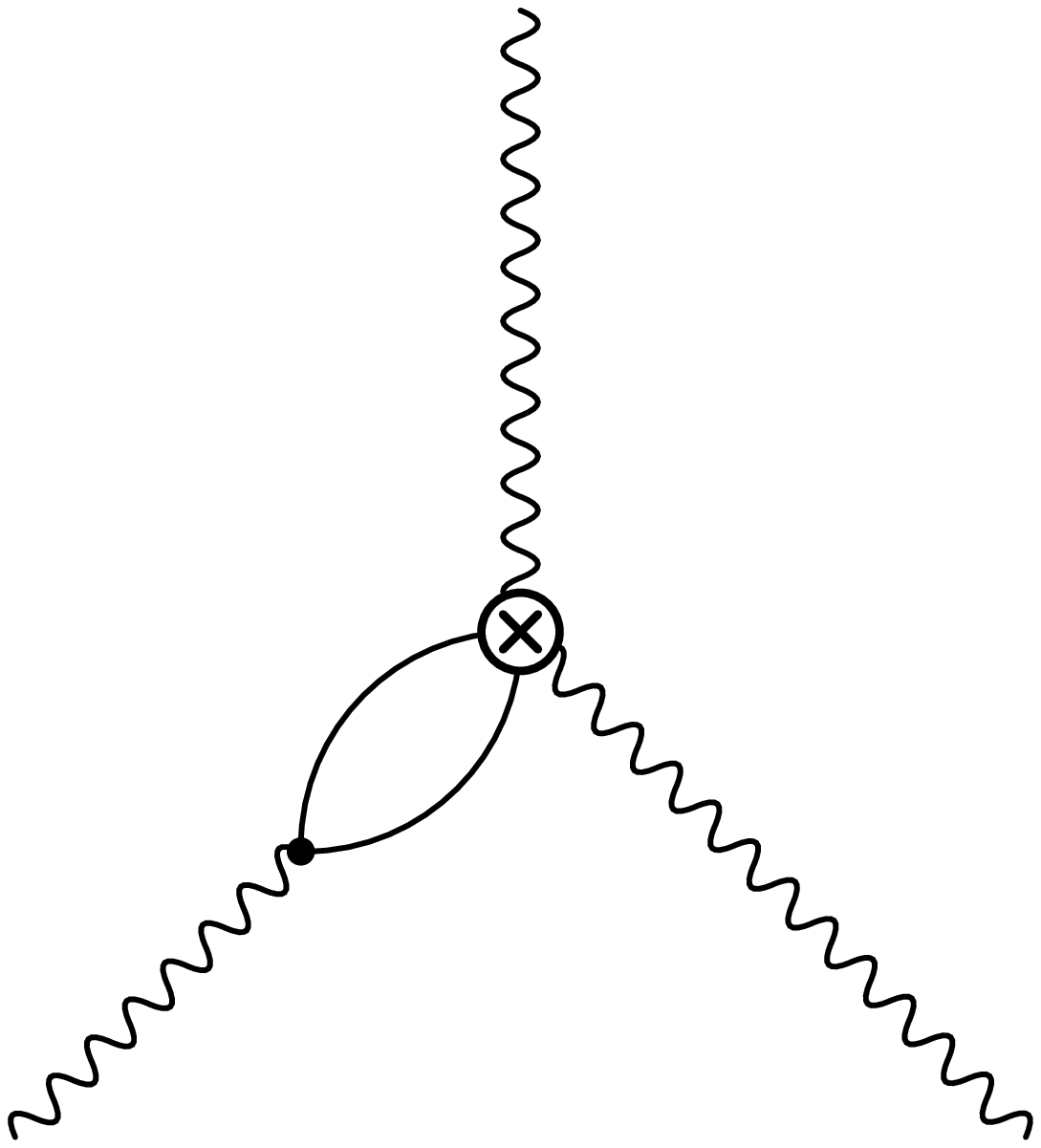,width=1.7cm} &
\leavevmode\psfig{file=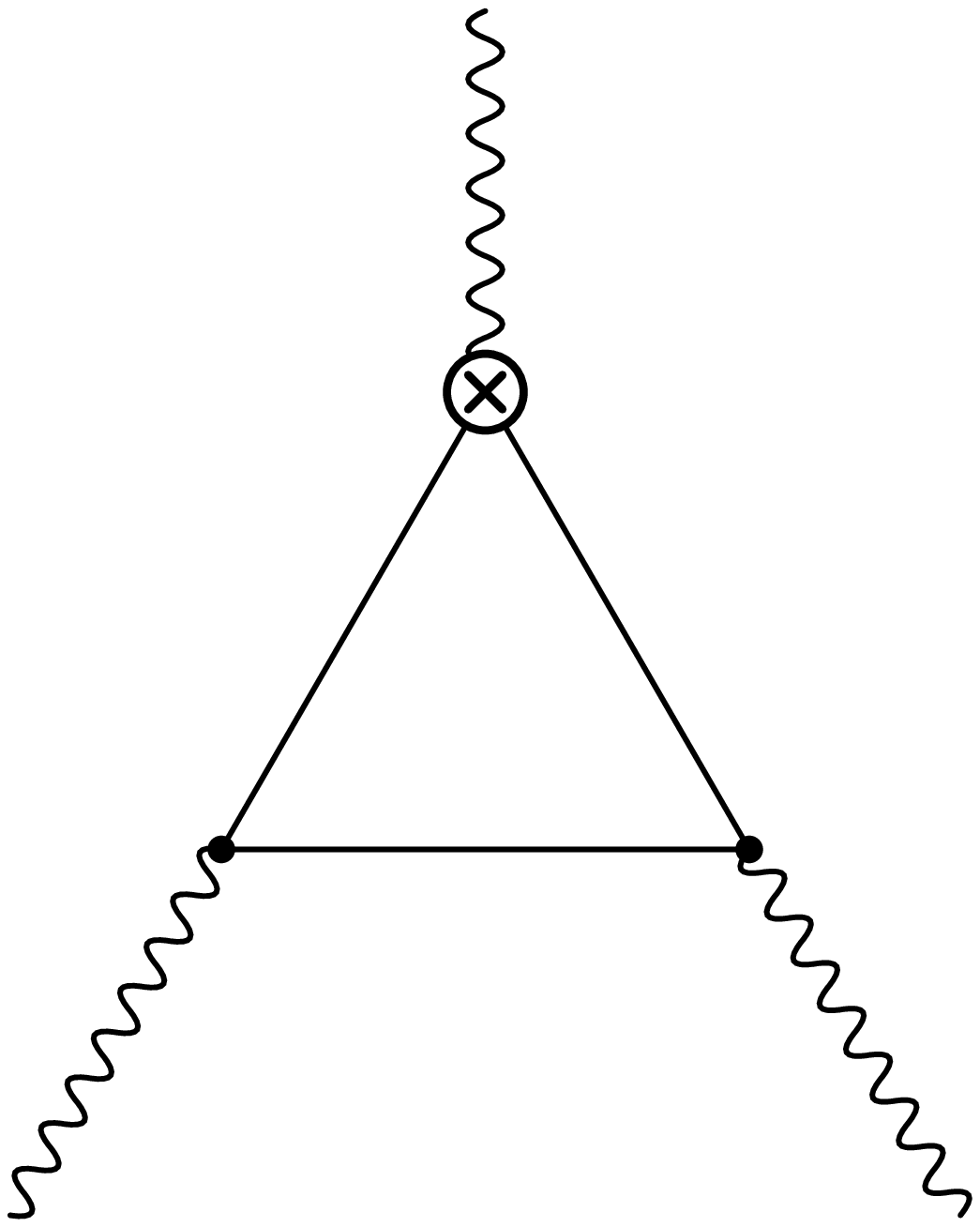,width=1.7cm}
\end{tabular}
\caption{1PI diagrams for three-point functions 
(full lines: quarks; wavy lines: gluons)}
\label{fig:DGM}
\end{center}
\end{figure}
\noindent
In the MS scheme, we express the renormalization constants
$Z_{ij}$ as,
\bea
Z_{ij} \equiv \delta_{ij} + 
\frac{g^{2}}{8 \pi^{2}(4-D)} X_{ij}
\;\;\;\;\;\;\; (i, j = 1, m, B, E_{F}, E_{G})
\label{eqn:7}
\eea
with $D$ the space-time dimension. The relevant components of $X_{ij}$ read:
\begin{eqnarray}
X_{11}&=& \frac{1}{3}C_{F} - 3C_{G} - \frac{2}{3}N_{f};\;\;\;\;\;
X_{1m} = \frac{2}{9}C_{F}; \;\;\;\;\; X_{1B} = -\frac{2}{9}N_{f},
\nonumber \\
X_{1E_{F}}&=& \frac{1}{6}C_{F}; \;\;\;\;\; 
X_{1E_{G}} = - \frac{2}{9}N_{f}; \;\;\;\;\;
X_{mm} = -6 C_{F},
\label{eq:xij}
\end{eqnarray}
where $C_{F}=\frac{N_{c}^{2}-1}{2N_{c}},~C_{G}=N_{c}$ for $N_{c}$ color,
and $N_{f}$ is the number of quark flavors.
The physically relevant component is only $X_{11}$. However it is
inevitable to identify other components to get a correct result for
$X_{11}$.

Now let us discuss the relation between our approach and
other ones based on the on-shell calculation. 
For this purpose, we note that eq.(\ref{eqn:6}) implies 
the following results for the 1PI three-point functions
including one-loop corrections (we consider the massless quark case):
\begin{eqnarray}
\Gamma_{gqq}^{\sigma}(k_1 , p, p') &=&
Z_{11}\left\langle R_{3,1}^{\sigma} \right\rangle_{gqq}^{(3)}
+ Z_{1 E_{F}}\left\langle R_{3, E}^{\sigma}\right\rangle_{gqq}^{(3)}
+ Z_{1 E_{G}}\left\langle T_{3, E}^{\sigma}\right\rangle_{gqq}^{(3)},
\label{eq:1piq} \\
\Gamma_{ggg}^{\sigma}(k_1 , k_2 , k_3 ) &=&
Z_{1B}\left\langle T_{3, B}^{\sigma}\right\rangle_{ggg}^{(3)}
+ Z_{1 E_{G}}\left\langle T_{3, E}^{\sigma}\right\rangle_{ggg}^{(3)},
\label{eq:1pig}
\end{eqnarray}
where $\Gamma_{gqq}^{\sigma}(k_1 , p,p')$ and
$\Gamma_{ggg}^{\sigma}(k_1 , k_2 , k_3 )$
are the Fourier transform of the 1PI Green's functions
$\Omega_{1}^{\mu} \langle 0|{\rm T} \Delta \cdot R_{3,1}^{\sigma}
A_{\mu} \psi \overline{\psi} |0 \rangle_{\rm 1PI}$
and $\Omega_{1}^{ \mu }\Omega_{2}^{ \nu }\Omega_{3}^{ \rho }
\langle 0 | {\rm T} \Delta \cdot R_{3,1}^{\sigma}
A_{ \mu } A_{ \nu } A_{ \rho } | 0 \rangle_{\rm 1PI}$
with $k_1 , p, p'(= - k_1 -p) , k_2 $ and $k_3 (= - k_1 - k_2 )$
the incoming (off-shell) external momenta.
The three-point vertices appearing in eqs.(\ref{eq:1piq}) and (\ref{eq:1pig})
are given by,
\begin{eqnarray}
\left\langle R_{3,1}^{\sigma}\right\rangle^{(3)}_{gqq} 
&=& 
	-\frac{1}{3}\left\langle T_{3,E}^{\sigma}\right\rangle^{(3)}_{gqq}
	=-\frac{g}{6} \hat{k}_1 
\left[ \gamma^{\sigma}, \slashl{\Omega_{1}}
\right] \gamma_{5} \slashl{\Delta}t^{a},
\label{eq:3pv1}\\
\left\langle R_{3,E}^{\sigma}\right\rangle^{(3)}_{gqq} 
&=& \frac{g}{3}\left(
   \hat{p}' \gamma^{\sigma} \slashl{\Omega_{1}}
     \gamma_5 \slashl{\Delta} - \hat{p}
   \slashl{\Omega_{1}} \gamma^{\sigma}\gamma_5 \slashl{\Delta} \right)
           t^a, \label{eq:3pv2}\\ 
\left\langle T_{3,B}^{\sigma}\right\rangle^{(3)}_{ggg} 
&=& - g f^{abc} \varepsilon^{\sigma \lambda \mu \nu}
\Delta_{\lambda} 
\Omega_{2 \mu} \Omega_{3 \nu}(k_1 \cdot \Omega_{1}) \hat{k}_1 
+\  (\mbox{cyclic perm's}),
\label{eq:3pv3}\\
\left\langle T_{3,E}^{\sigma}\right\rangle^{(3)}_{ggg} 
&=& - g f^{abc} \varepsilon^{\sigma \lambda \mu \nu}
\Delta_{\lambda} 
\left( \Omega_{2 \mu} \Omega_{3 \nu}(k_1 \cdot \Omega_{1})
\hat{k}_1 + k_{1 \mu} \Omega_{1 \nu} (\Omega_{2} \cdot \Omega_{3})
(\hat{k}_3 - \hat{k}_2 ) \right)
\nonumber \\
& & \qquad +\ (\mbox{cyclic perm's})-
\left\langle T_{3,B}^{\sigma}\right\rangle^{(3)}_{ggg},
\label{eq:3pv4}
\end{eqnarray}
where $\hat{p}\equiv\Delta\cdot p$, etc., and ``(cyclic perm's)'' denotes
the terms due to the cyclic permutation of the labels for the external lines
$(k_1 , \Omega_{1}, a), (k_2 , \Omega_{2}, b)$,
and $(k_3 , \Omega_{3}, c)$. We now take the on-shell limit of the
above expressions.
We can identify the vectors $\Omega_{i}^{\mu}$ to be the polarization vectors
$\epsilon^{\mu}_i (k_i )$ of the corresponding gluons.
Therefore the on-shell limit is realized by taking  
$\slashs{p} = \slashs{p}' = k_1^{2} = k_1 \cdot \Omega_{1} = 0$ for 
eq.(\ref{eq:1piq}), and
$k_1^{2}= k_2^{2} = k_3^{2} = k_1 \cdot \Omega_{1}
=k_2 \cdot \Omega_{2} = k_3 \cdot \Omega_{3} = 0$ for eq.(\ref{eq:1pig}).
We see $\left\langle T_{3, B}^{\sigma} \right\rangle_{ggg}^{(3)} =0$
from eq.(\ref{eq:3pv3}) and thus the BRST-exact operator does not
contribute in this limit. On the other hand, we find that 
$\left\langle R_{3,E}^{\sigma} \right\rangle_{gqq}^{(3)}$,
$\left\langle T_{3,E}^{\sigma} \right\rangle_{gqq}^{(3)}$, and
$\left\langle T_{3,E}^{\sigma} \right\rangle_{ggg}^{(3)}$
of (\ref{eq:3pv1}), (\ref{eq:3pv2}), and (\ref{eq:3pv4})
do not vanish; the EOM operators do contribute
to the three-point functions eqs.(\ref{eq:1piq}) and (\ref{eq:1pig})
even in the on-shell limit.

There exist, however, the one-particle-reducible (1PR)
contributions of the same order as the 1PI ones (see Fig.2 ).
\begin{figure}[H]
\begin{center}
\begin{tabular}{cccccc}
\leavevmode\psfig{file=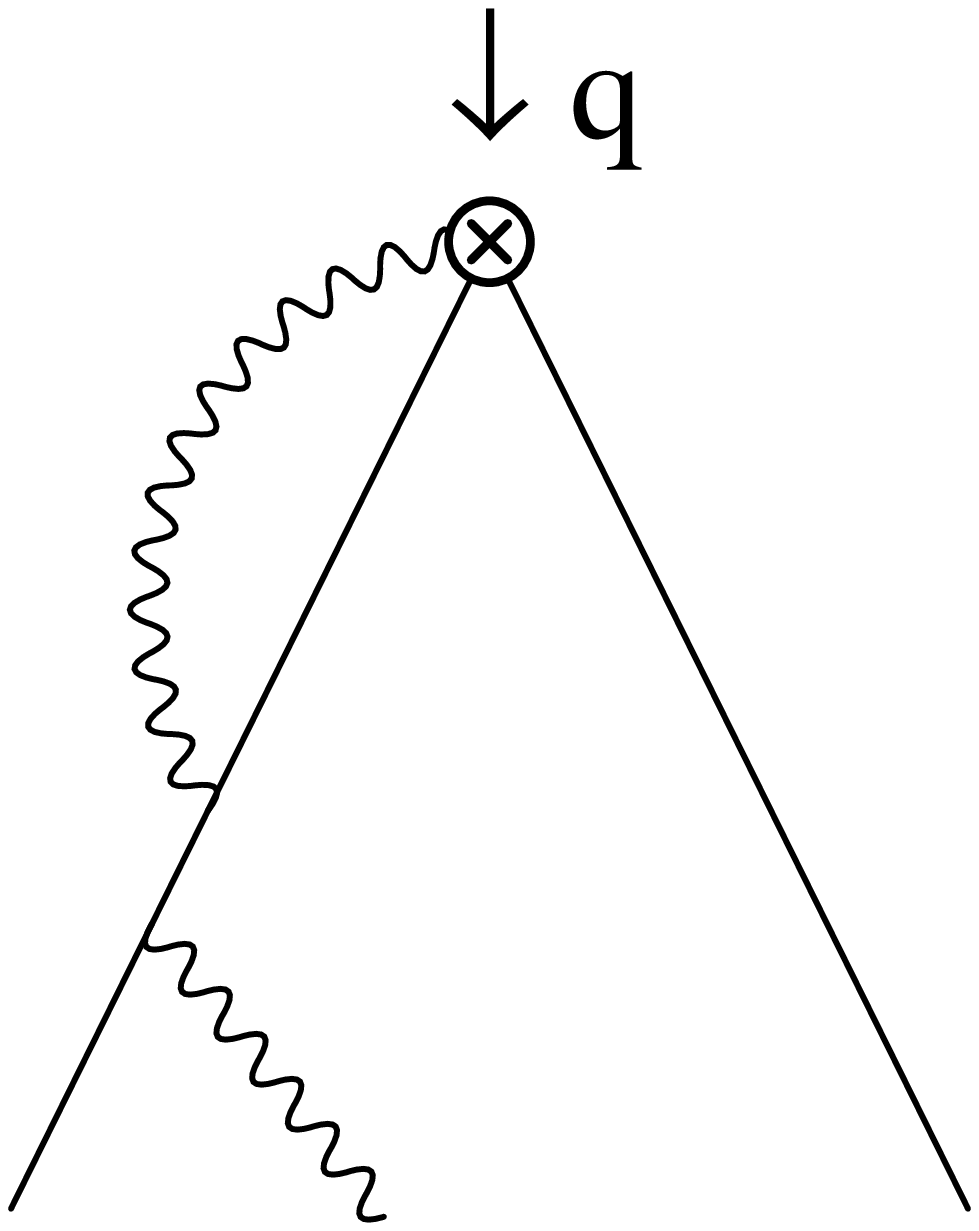,width=1.7cm} &
\leavevmode\psfig{file=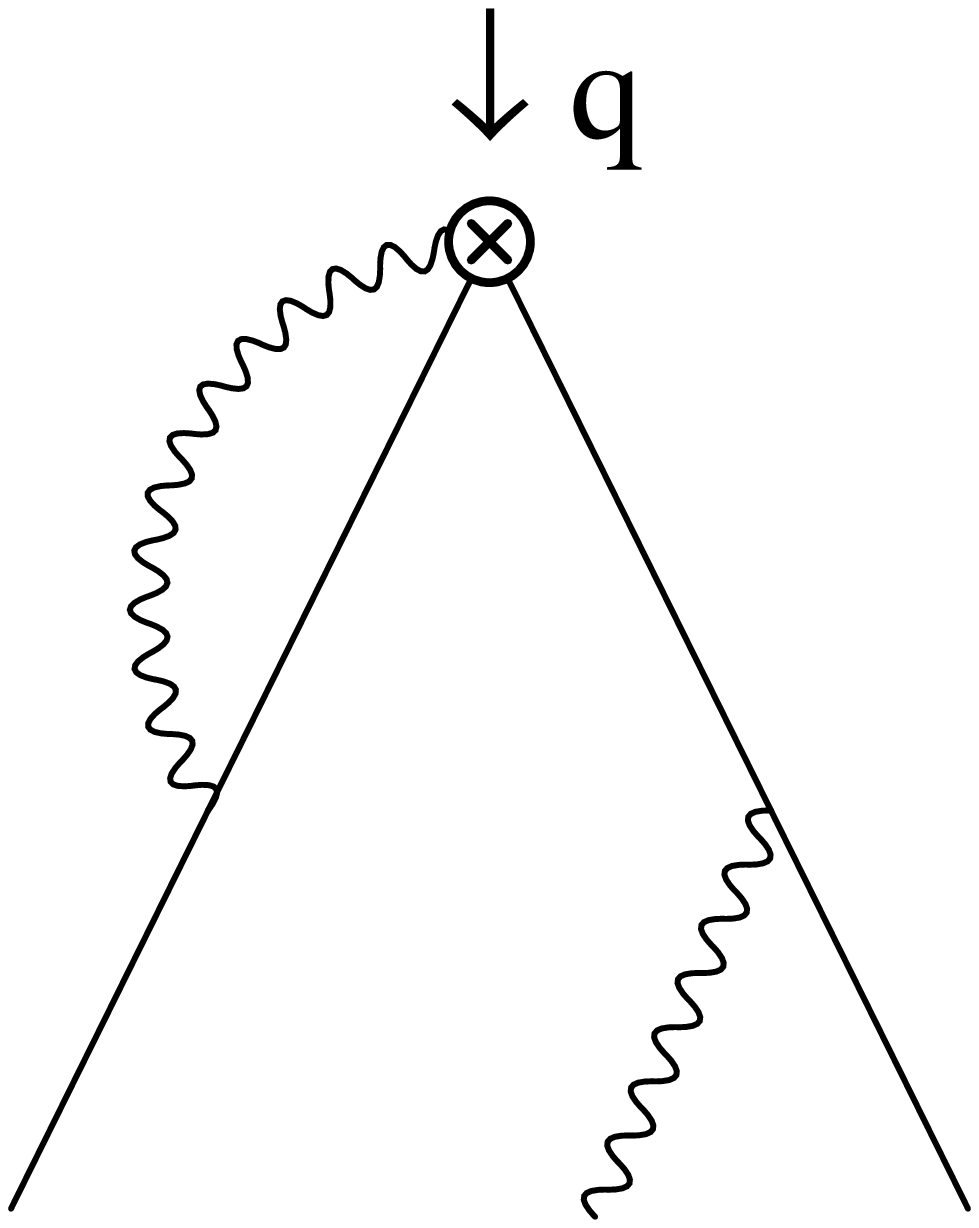,width=1.7cm} &
\leavevmode\psfig{file=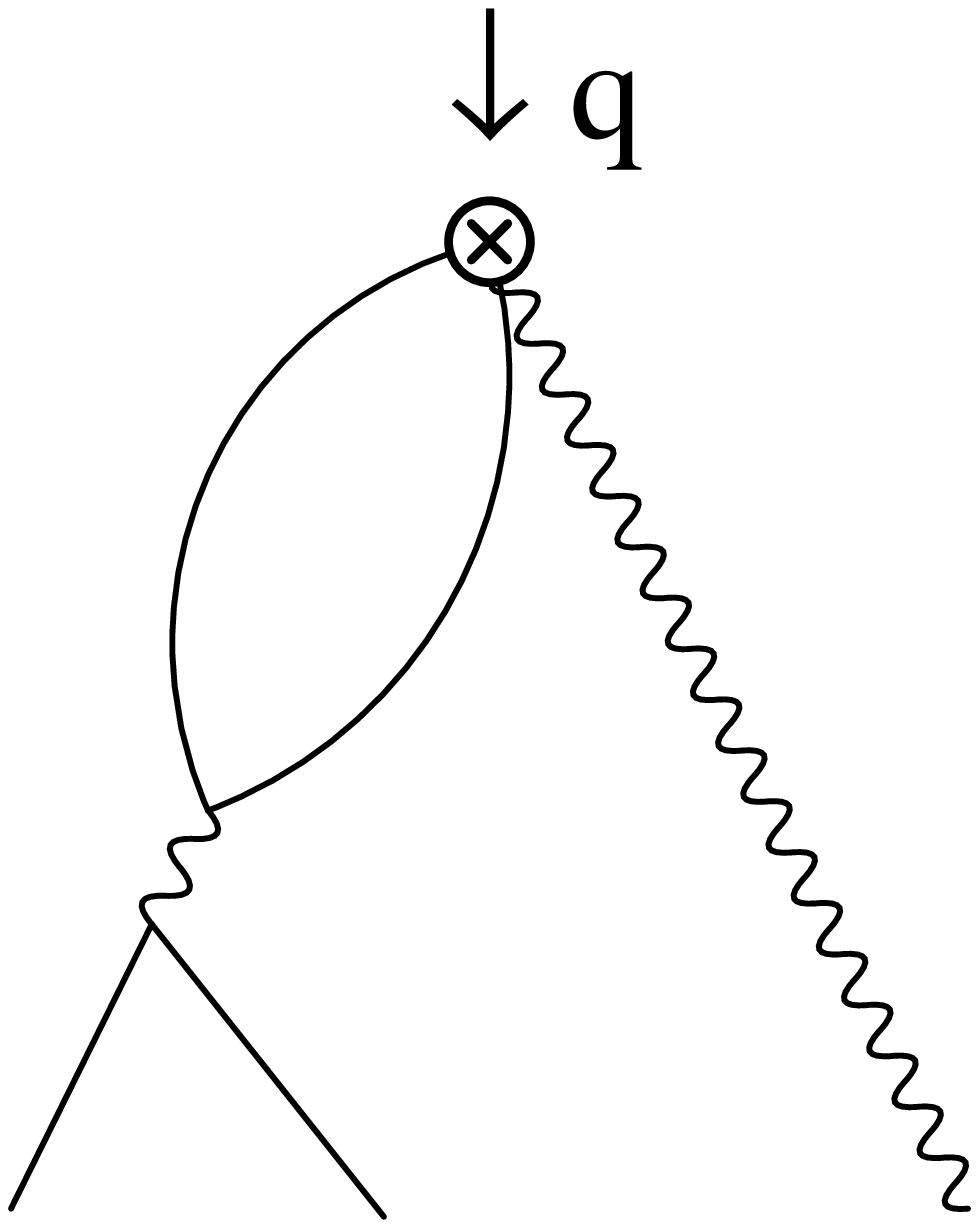,width=1.7cm} &
\leavevmode\psfig{file=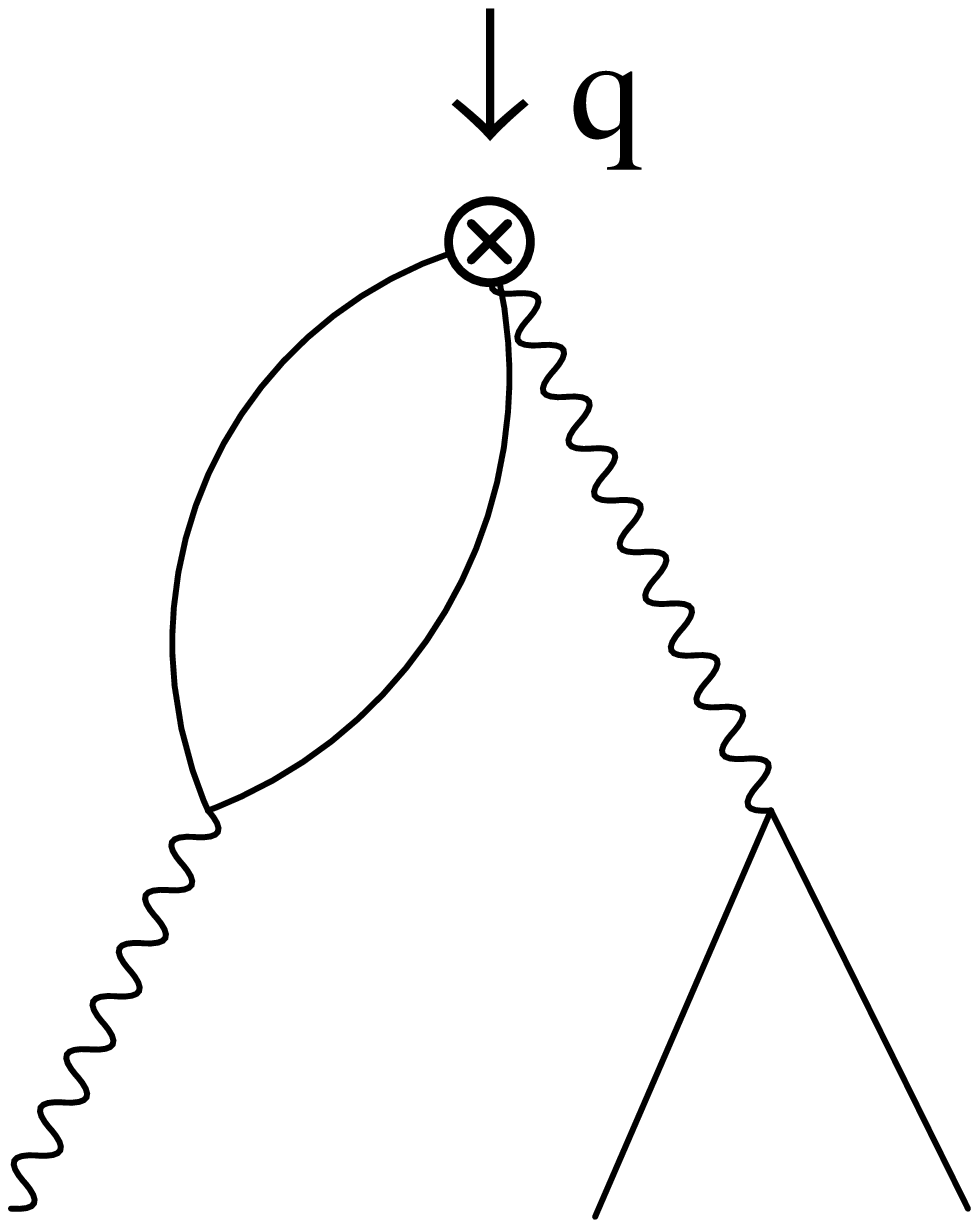,width=1.7cm} &
\leavevmode\psfig{file=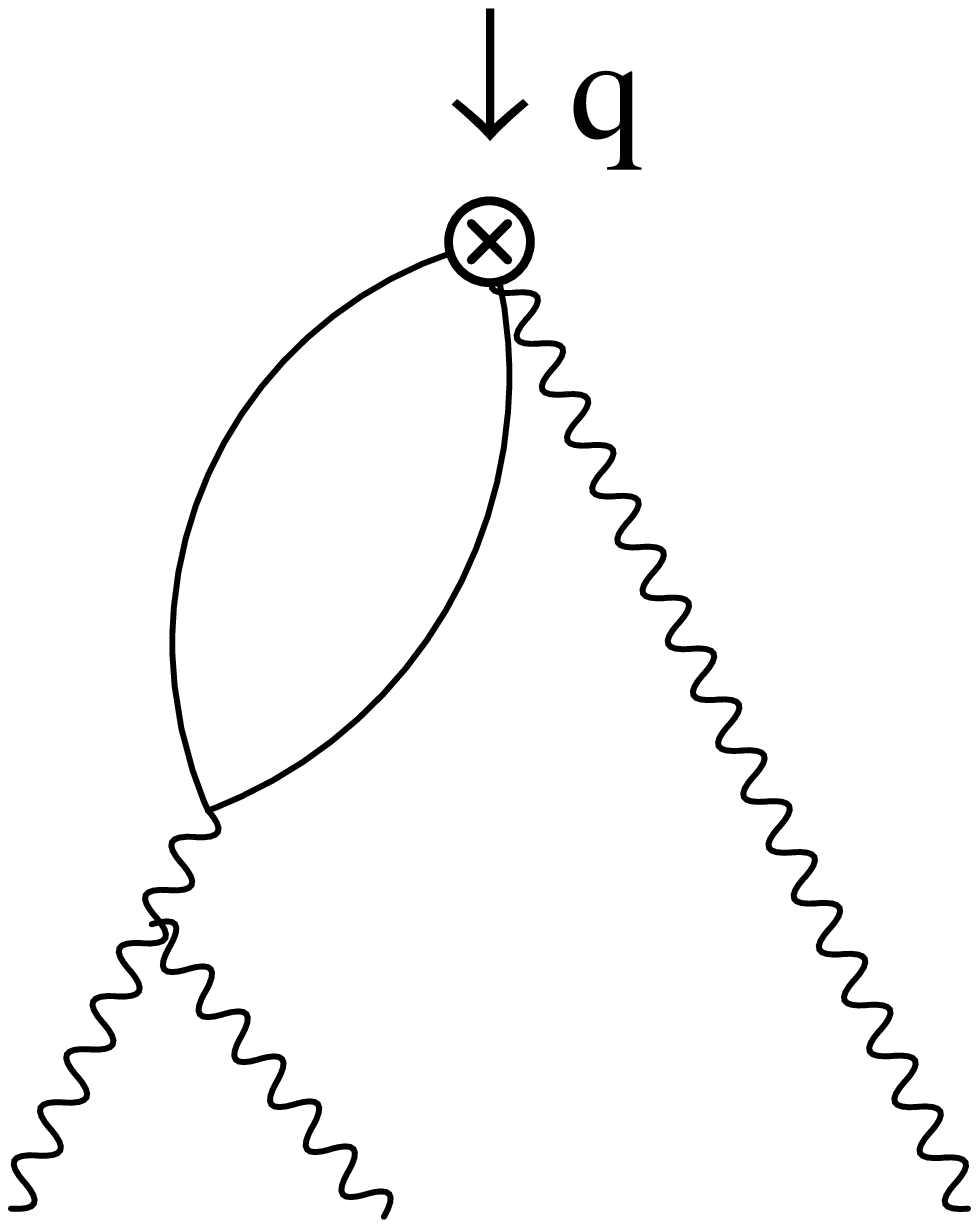,width=1.7cm} &
\leavevmode\psfig{file=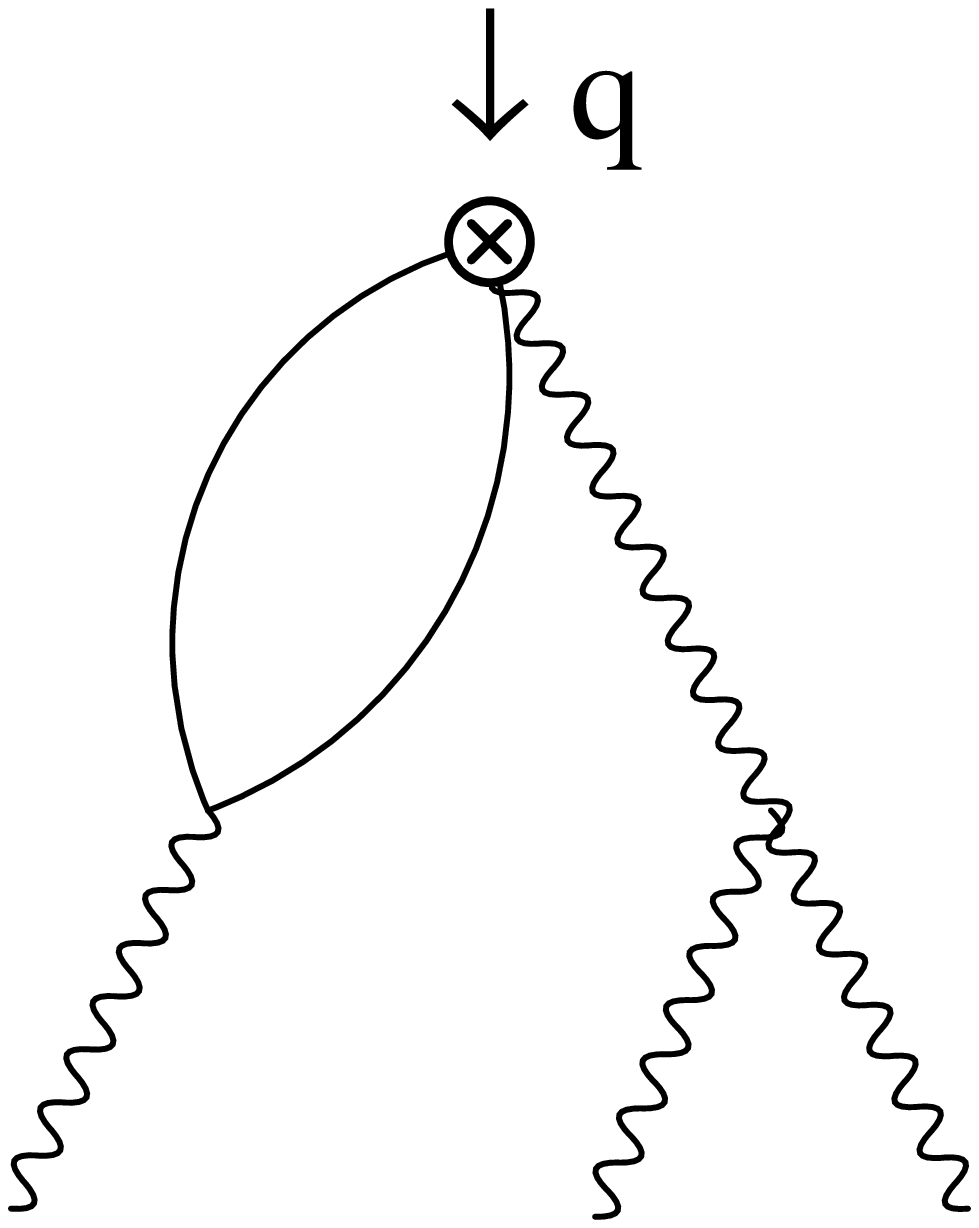,width=1.7cm}
\end{tabular}
\caption{1PR diagrams for three-point functions 
(full lines: quarks; wavy lines: gluons)}
\label{fig:DGM2}
\end{center}
\end{figure}
\noindent
We denote the 1PR contributions corresponding to
eqs.(\ref{eq:1piq}) and (\ref{eq:1pig}) as
$\widetilde{\Gamma}^{\sigma}_{gqq}(k_1 , p, p')$
and $\widetilde{\Gamma}^{\sigma}_{ggg}(k_1 , k_2 , k_3 )$.
Here it should be noted that the on-shell limit of three massless
particles corresponds to a very singular (collinear) configuration in the
momentum space. Actually the on-shell limit of the 1PR diagram (Fig.2)
becomes indefinite due to the collinear singularity.
This technical difficulty can be avoided by calculating
the \lq\lq non-forward\rq\rq\  matrix element of the
composite operator \cite{JCOL2}.
Namely we calculate the diagram with the composite
operators at nonzero momentum
($-q \equiv k_1 + p + p' \equiv k_1 + k_2 + k_3 \neq 0$ ).  
Then we \lq\lq define\rq\rq\   $\widetilde{\Gamma}^{\sigma}_{gqq}$ and
$\widetilde{\Gamma}^{\sigma}_{ggg}$ to be the $q \to 0$
limit of the matrix element after putting the external fields
on the mass shell.

The contributions from Fig.2 are given by,
\begin{eqnarray}
\widetilde{\Gamma}^{\sigma}_{gqq}(k_1 , p, p')
   &=& - g \slashl{\Omega}_{1} t^{a} \frac{1}{\slashs{p} + \slashs{q}}
       \Pi_{qq}^{\sigma}(p , q) + \Pi_{qq}^{\sigma}(- p'- q , q)
    \frac{1}{\slashs{p'}+\slashs{q}} g \slashl{\Omega}_{1} t^{a}
    \nonumber\\
  & & \qquad\qquad +\  \Omega_{1\mu}
       \Pi_{gg}^{\sigma \mu \nu}(k_1 , q) \frac{1}{(k_1 + q)^{2}}
          g\gamma_{\nu} t^{a}, \label{eq:1pr1} \\
\widetilde{\Gamma}^{\sigma}_{ggg}(k_1 , k_2 , k_3 )
   &=& \Omega_{1\mu} \Pi_{gg}^{\sigma \mu \nu}(k_1 , q)
          \frac{-i}{(k_1 + q)^{2}}
   gf^{abc}V_{\nu \alpha \beta}(- k_1 - q , k_2 , k_3 ) \Omega_{2}^{\alpha}
      \Omega_{3}^{\beta} \label{eq:1pr2}\\
   & & \qquad\qquad + \ (\mbox{cyclic perm's}) \nonumber,
\end{eqnarray}
where $V_{\mu_{1}\mu_{2}\mu_{3}}(p_{1}, p_{2}, p_{3})
= (p_{1}- p_{2})_{\mu_{3}}g_{\mu_{1}\mu_{2}} +
(\mbox{cyclic perm's})$ is the usual three-gluon vertex.
$\Pi_{qq}^{\sigma}(p , q)$ and 
$\Pi_{gg}^{\sigma\mu \nu}(k , q)\delta^{ab}$ correspond to
$\langle 0|{\rm T} \Delta \cdot R_{3,1}^{\sigma}(q) 
\psi(-p-q) \overline{\psi}(p)|0 \rangle_{\rm 1PI}$
and 
$\langle 0 | {\rm T} \Delta \cdot R_{3,1}^{\sigma}(q)
A^{a \mu}(k ) A^{b \nu}(-k - q)| 0 \rangle_{\rm 1PI}$
with the off-shell external momenta.
Their explicit expressions (divergent term) read,
\begin{eqnarray}
  \Pi_{qq}^{\sigma}(p,q)
   &=& Z_{1(q)} \frac{1}{3}  \left( (\hat{p} + \hat{q})
       (\slashs{p} + \slashs{q}) \gamma_{5}
   \gamma^{\sigma}\slashl{\Delta} + \hat{p} \gamma_{5}\gamma^{\sigma}
    \slashl{\Delta}\slashs{p} \right),
      \label{eq:2pv1}\\
  \Pi_{gg}^{\sigma \mu \nu}(k,q) &=& Z_{2(q)}\ 
   (-i)\  \varepsilon^{\sigma \lambda \alpha \beta} \Delta_{\lambda}
  \Biggl[ k_{\alpha}
     g_{\beta}^{\mu} \left\{ (k+q)^{2}\Delta^{\nu}
     - (\hat{k}+ \hat{q}) (k^{\nu} + q^{\nu}) \right\}
          \nonumber\\
     & & \qquad\qquad\qquad\qquad  - (k + q)_{\alpha} g_{\beta}^{\nu}
       \left( k^{2} \Delta^{\mu} - \hat{k}k^{\mu} \right) \Biggr] .
   \label{eq:2pv3}
\end{eqnarray}
where
\[ Z_{1(q)} = Z_{1{E_F}} \ , \qquad Z_{2(q)} = Z_{1{E_G}}. \]
Substituting eqs.(\ref{eq:2pv1})-(\ref{eq:2pv3}) into
eqs.(\ref{eq:1pr1}) and (\ref{eq:1pr2}),
we can go over to the on-shell limit with $q$ being kept non-zero.
After this procedure, we take the $q \to 0$ limit leading to
$\widetilde{\Gamma}_{gqq}^{\sigma} = - Z_{1E_{F}} \left\langle
R_{3,E}^{\sigma} \right \rangle_{gqq}^{(3)} - Z_{1E_{G}} 
\left\langle
T_{3,E}^{\sigma} \right \rangle_{gqq}^{(3)}$
and 
$\widetilde{\Gamma}_{ggg}^{\sigma} = - Z_{1E_{G}} 
\left\langle
T_{3,E}^{\sigma} \right \rangle_{ggg}^{(3)}$.
Therefore, the nonzero terms coming from the EOM operators
exactly cancel out in the sum of the 1PI and the 1PR contributions
for the on-shell external momenta:
\begin{equation} 
\Gamma_{gqq}^{\sigma} + \widetilde{\Gamma}_{gqq}^{\sigma}
= Z_{11}\left\langle R_{3,1}^{\sigma} \right\rangle_{gqq}^{(3)}\ , \qquad
\Gamma_{ggg}^{\sigma} + \widetilde{\Gamma}_{ggg}^{\sigma}=0.
\label{eq:result}
\end{equation}
The results eq.(\ref{eq:result}) correspond to the on-shell calculations
of the evolution kernel in the literature\cite{BKL},
where neither the EOM operators nor the alien
operators appear. The non-zero contributions from the EOM operators in 
the on-shell limit and their cancellation in the sums eq.(\ref{eq:result})
are novel features in the three-point functions, whose
consideration is indispensable for the twist-3 operators.
This phenomenon is consistent with the well-known theorem that
the physical matrix elements of the EOM operators vanish\cite{JCOL1}, 
because both 1PI and 1PR diagrams contribute to the physical
matrix elements. We emphasize that, in our demonstration,
a calculation of the \lq\lq physical\rq\rq\ matrix element
of the composite operators, in general, stays with some
subtleties and dangers due to infrared singularities.
One of the consistent method to avoid this problem from the
beginning might be that all diagrams (1PI as well as 1PR) are calculated with
the composite operators at nonzero momentum\cite{JCOL2}. 
In this case, however, we must consider additional composite operators which
differ from the previous ones by the total derivative and this fact brings in
another complexity\footnote{The operator $R^{\sigma}_{3,1}$
is unique even if $q \neq 0$.}.
Therefore the off-shell Green's functions are much more
tractable and easy to calculate especially for the general
moment $n$.

Finally, we mention the $Q^{2}$-evolution obtained from 
eqs.(\ref{eqn:6})-(\ref{eq:xij}): If we neglect the quark mass operator 
$\Delta \cdot R_{3,m}^{\sigma}$, only one
operator $\Delta \cdot R_{3,1}^{ \sigma }$ contributes to the
$Q^{2}$-evolution of the physical matrix elements.
The relevant renormalization constant is $Z_{11}$, which gives 
\bea
\int_{0}^{1}dx x^{2}g_{2}^{tw.3}(x, Q^{2})
~=~
\left(
\frac{\alpha_{s}(Q^{2})}{\alpha_{s}(\mu^{2})}
\right)^{(3C_{G} - C_{F}/3 + 2N_{f}/3)/b}
\int_{0}^{1}dx x^{2}g_{2}^{tw.3}(x, \mu^{2}). 
\label{eqn:13}
\eea
Here $b=\frac{11}{3}N_{c}-\frac{2}{3} N_{f}$, 
and $\alpha_{s}(Q^{2})$ is the running coupling constant.
$g_{2}^{tw.3}(x,{\mu}^2)$ is the twist-3 contribution of $g_{2}(x,{\mu}^{2})$
after subtracting out the Wandzura-Wilczek piece\cite{JA},
and its moment is equal to the nucleon matrix element of 
$\Delta \cdot R_{3,1}^{\sigma}(\mu^{2})$ up to irrelevant factor.
The result eq.(\ref{eqn:13}) coincides with that of Refs.\cite{SV,BKL}, 
though our approach is quite different from those works;
this fact confirms the theoretical prediction eq.(\ref{eqn:13}), and
also demonstrates the efficiency of our method. We note that 
the term $\frac{2}{3} N_{f}$ would be absent from the exponent of
eq.(\ref{eqn:13}) if we consider the non-singlet case\cite{KOD1,KOD2}.
For $N_{c}=3$ and $N_{f}=4$, the exponent of eq.(\ref{eqn:13}) is
$\frac{101}{9b}$, while the corresponding exponent for the non-singlet case
is $\frac{77}{9b}$. Thus the singlet channel obeys rather stronger
$Q^{2}$-evolution compared to the non-singlet one.

In the present study, we have developed a manifestly covariant approach
to investigate the flavor singlet part of $g_{2}(x,Q^{2})$.
We derived the operator identities which relate the two-particle operators
with the three-particle ones. We have chosen the three-particle
operators as an independent operator's basis.
To identify the renormalization constants correctly,
the off-shell Green's functions are considered.
In this case, it is indispensable to consider the mixing of the gauge
invariant operators with the EOM as well as the BRST-exact operators.
As an application, we investigated in detail the
renormalization for the lowest ($n=3$) moment. 
We made contact with other methods\cite{BKL}, and also confirmed
the prediction of the $Q^{2}$-evolution obtained by them.
The off-shell Green's functions are free from the infrared singularity
coming from the collinear configuration as well as any
other unphysical singularities. We believe that calculating
the off-shell Green's functions is the safest and the most
straightforward method to obtain the anomalous dimensions of the
higher-twist operators.

\vspace{2cm}

We would like to thank T.Uematsu for valuable discussions
and suggestions.
Y. Y would like to thank L. Trueman and S. Ohta for useful discussions.


\end{document}